\documentclass[
    ,final            % use final for the camera ready runs
  ]
  {aipproc}

\layoutstyle{6x9}

\usepackage{txfonts}
\usepackage{natbib}
\bibpunct{(}{)}{;}{a}{}{,} % to follow the A&A style

\newcommand\T{\rule{0pt}{2.0ex}}       % Top VSpace inside tables         
\def\mso{\,\mathrm{M}_\odot}

 \def\kms{\, {\rm km}\, {\rm s}^{-1}}

 \def\simle{\mathrel{\hbox{\rlap{\hbox{\lower4pt\hbox{$\sim$}}}\hbox{$<$}}}}
 \def\simgr{\mathrel{\hbox{\rlap{\hbox{\lower4pt\hbox{$\sim$}}}\hbox{$>$}}}}

 \def\vk{\varv_{\,\mathrm{K}}} 
 \def\vik{\varv_{\,\mathrm{init}}/\vk}

\begin{document}

\title{Long GRBs from binary stars: runaway, Wolf-Rayet progenitors}


\author{M.Cantiello}{
  address={Astronomical Institute, Utrecht University,\\
              Princetonplein 5, 3584 CC, Utrecht, The Netherlands}
}

\author{S.-C.Yoon}{
  address={Department of Astronomy and Astrophysics, University of California, Santa Cruz, CA 95064, USA}
}

\author{N.Langer}{
  address={Astronomical Institute, Utrecht University,\\
              Princetonplein 5, 3584 CC, Utrecht, The Netherlands}
}
\author{M.Livio}{
  address={Space Telescope Science Institute, 3700 San Martin Drive, Baltimore, MD 21218 }
}

\begin{abstract}
  The collapsar model for long gamma-ray bursts requires a rapidly rotating
  Wolf-Rayet star as progenitor.
  We test the idea of producing rapidly rotating Wolf-Rayet stars 
  in massive close binaries through mass accretion and consecutive
  quasi-chemically homogeneous evolution --- the latter had previously been shown
  to provide collapsars below a certain metallicity threshold for single stars.
  The binary channel presented here may provide a means
  for massive stars to obtain the high rotation rates required to evolve
  quasi-chemically homogeneous and fulfill the collapsar scenario.
  Moreover, it suggests that a possibly large fraction of long gamma-ray bursts
  occurs in runaway stars.
\end{abstract}

\maketitle

%%%%%%%%%%%%%%%%%%%%%%%%%%%%%%%%%%%%%%%%%%%%
%% MAINMATTER
%%%%%%%%%%%%%%%%%%%%%%%%%%%%%%%%%%%%%%%%%%%%

\section{Introduction}
Long gamma-ray bursts are thought to be produced by a subset of dying massive
and possibly metal-poor stars \citep{2005MNRAS.362..245J,2006ApJ...638L..63L,2007Modjaz}.
Within the currently favored collapsar scenario \citep{1993ApJ...405..273W},
the burst is produced by a rapidly rotating massive Wolf-Rayet (WR) star
whose core collapses into a black hole \citep{1999ApJ...524..262M}. 
While single star evolution models without internal magnetic fields
can produce such configurations \citep{2005A&A...435..247P,2005A&A...443..581H},
only models including magnetic fields are capable of reproducing the slow
spins of young Galactic neutron stars \citep{2005ApJ...626..350H,2006ApJS..164..130O}
and white dwarfs \citep{2007Suijs}, due to the magnetic core-envelope
coupling during the giant stage. 
  
\citet{2005A&A...443..643Y}, \citet{2006A&A...460..199Y} and \citet{2006ApJ...637..914W}
recently showed that below a certain metallicity threshold, very rapidly rotating single stars
avoid the magnetic braking of the core through the so-called quasi-chemically homogeneous
evolution: rotationally induced mixing processes keep the star close to chemical
homogeneity, and thus the giant stage is avoided altogether. 
While these models are successful in producing models which fulfill
all constraints of the collapsar model, they require very rapid initial
rotation.

The question thus arises whether the quasi-chemically homogeneous evolution of 
massive stars can also be obtained in mass transferring massive binary systems \citep{1994A&A...290..129V},
since in such systems the mass gainer can be spun-up to close to critical
rotation \citep[see][]{2005A&A...435.1013P,2005A&A...435..247P}, independent of its initial rotation rate.

\section{The Model}

We use a 1-D hydrodynamic binary evolution code to simulate the evolution of a 
  16+15 $\mso$ binary model with an initial orbital period of 5 days and SMC metallicity (Z=0.004).  
  Internal differential rotation, rotationally induced mixing and magnetic fields are included 
  in both components, as well as non-conservative mass and angular momentum transfer, and tidal
  spin-orbit coupling. Detailed description of the code and of the adopted physics can be found in \citet{cyl+07} and references therein.
 
We chose an early Case~B system with an initial
mass ratio close to one for two reasons. Firstly, the expected mass transfer efficiency
for this case was about 60\% (meaning that 60\% of the transfered matter can
be retained by the mass gainer), based on the calculations by \citet{2001WellsteinPhd},
\citet{2004IAUS..215..535.}, and \citet{2005A&A...435.1013P}.
Secondly, a Case~B rather than Case~A system was chosen to avoid synchronization
{\em after} the major mass transfer phase. 

 \begin{figure}
 \begin{minipage}{8.15cm}
  \resizebox{\hsize}{!}{\includegraphics{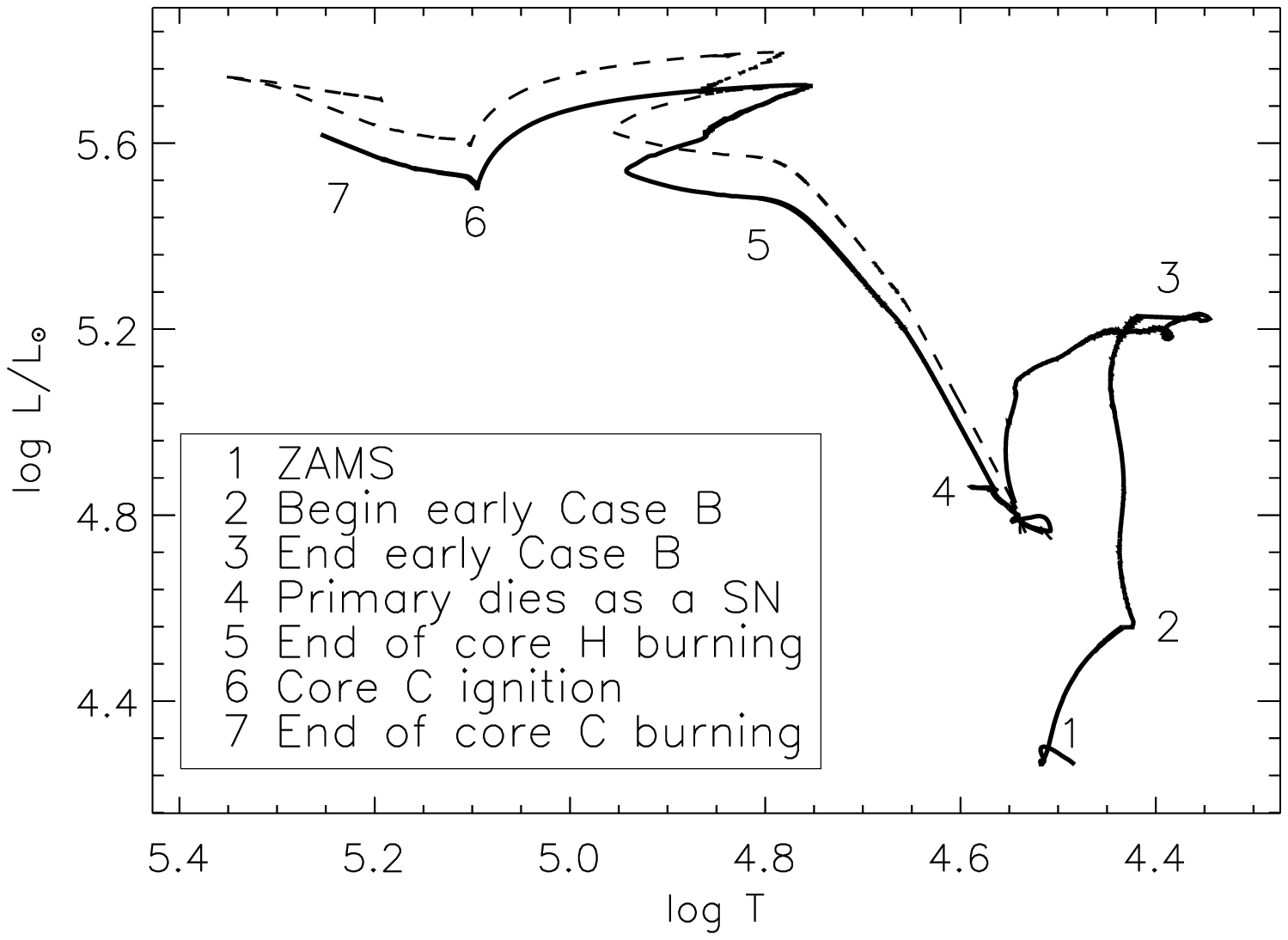}}
 \end{minipage}
 \hfill
 \begin{minipage}{6.95cm} 
   \resizebox{\hsize}{!}{\includegraphics{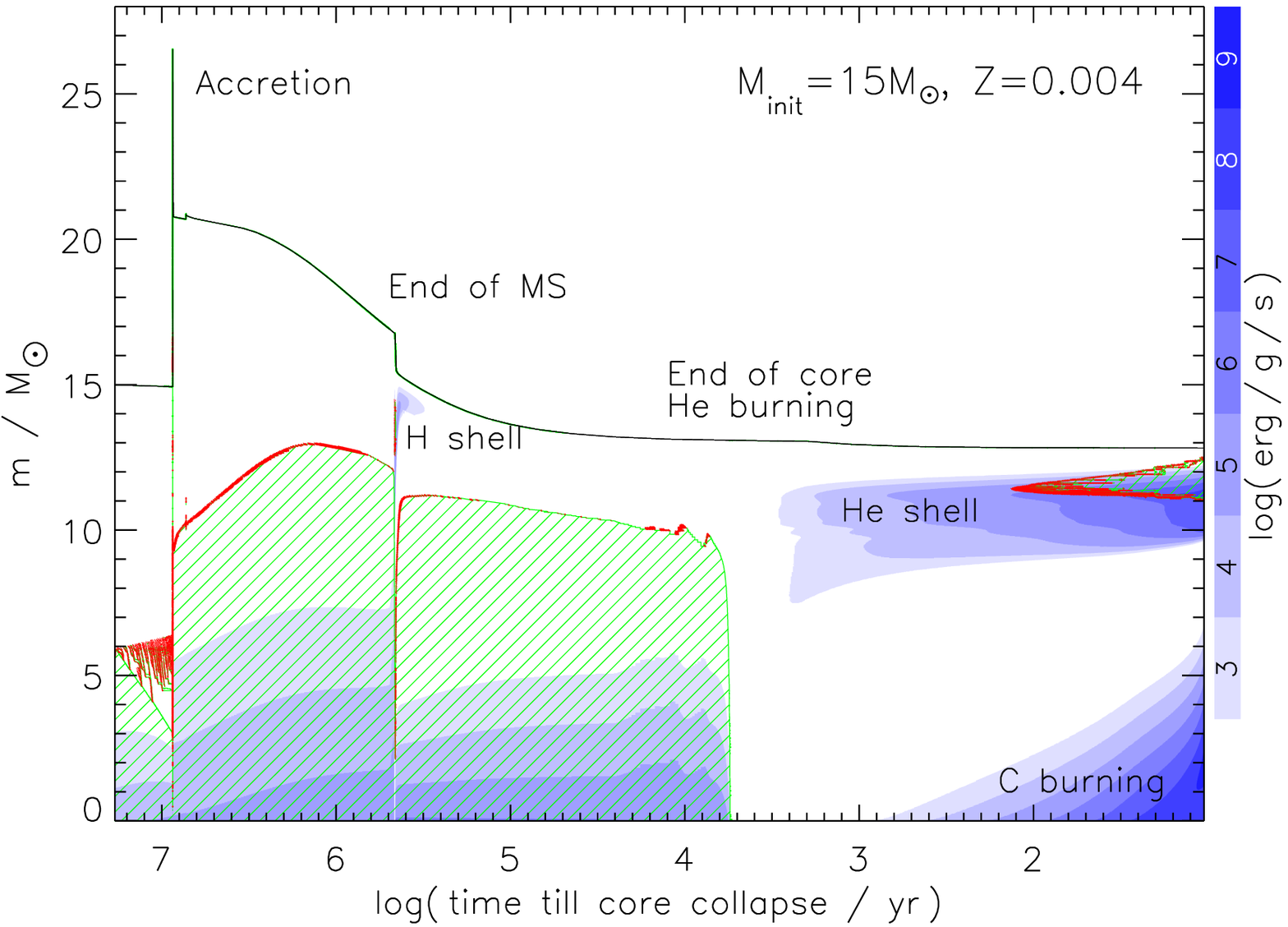}}
 \end{minipage}
\caption{{\bf Left panel:} evolutionary track of the mass gainer in our $16\mso + 15\mso$ 
early Case~B binary model (5~d initial orbital period) in the HR diagram (solid line),
from the zero age main sequence up to core carbon exhaustion. 
The main evolutionary phases are labeled by numbers (see legend). 
The dashed line shows the evolutionary track of a very rapidly rotating 
($\vik=0.9$) $24\mso$ single star.
Both stars have SMC metallicity, and undergo quasi-chemically homogeneous evolution
(see text). {\bf Right panel:} evolution of the internal structure of the mass gainer of the computed $16\mso + 15\mso$
early Case~B binary sequence, as function of time, 
from the zero-age main sequence to core carbon exhaustion. The time axis is logarithmic,
with the time of core collapse as zero point.
Convective layers are hatched. Semiconvective layers are marked by dots (red dots in the electronic version). 
Gray (blue) shading indicates nuclear energy generation (color bar to the right of the figure).
The topmost solid line denotes the surface of the star. }
\label{plots}
 \end{figure}

%\begin{figure}
%  \includegraphics[height=.3\textheight]{7115fig1}
%  \caption{Evolutionary track of the mass gainer in our $16\mso + 15\mso$ 
%early Case~B binary model (5~d initial orbital period) in the HR diagram (solid line),
%from the zero age main sequence up to core carbon exhaustion. 
%The main evolutionary phases are labeled by numbers (see legend). 
%The dashed line shows the evolutionary track of a very rapidly rotating 
%($\vik=0.9$) $24\mso$ single star.
%Both stars have SMC metallicity, and undergo quasi-chemically homogeneous evolution
%(see text).}
% \label{hrd}
%\end{figure}

%\begin{figure}
%  \includegraphics[height=.3\textheight]{7115fig3}
%  \caption{Evolution of the internal structure of the mass gainer of the computed $16\mso + 15\mso$
%  early Case~B binary sequence, as function of time, 
% from the zero-age main sequence to core carbon exhaustion. The time axis is logarithmic,
% with the time of core collapse as zero point.
% Convective layers are hatched. Semiconvective layers are marked by dots (red dots in the electronic version). 
% Gray (blue) shading indicates nuclear energy generation (color bar to the right of the figure).
% The topmost solid line denotes the surface of the star.}
% \label{kipp_bin}
%\end{figure}

%%%%%%%%%%%%%%%%%%%%%%%%%%%%%%%%%%%%%%%%%%%%
%% SAMPLE TABLE
%%
%% Shows the use of \tablehead and \tablenote
%% macros
%%%%%%%%%%%%%%%%%%%%%%%%%%%%%%%%%%%%%%%%%%%%

\begin{table}[tbp]
\caption{Major evolutionary phases of the computed $16\mso + 15\mso$ early Case~B binary sequence. 
The binary calculation ends after core carbon
exhaustion of the mass loser (the primary), 
and the mass gainer (the secondary) is then evolved as a single star. 
We show evolutionary time, masses of both stars, orbital period,
surface rotational velocity of both stars, surface and core helium mass fraction
of the mass gainer, and orbital velocity of the mass gainer.
The abbreviations for the evolutionary phases are:
ZAMS = zero age main sequence; 
ECHB = end core hydrogen burning; ICB= ignition of carbon burning; ECCB = end core carbon burning.
The numbered evolutionary stages correspond to those given in Fig.~\ref{plots}, left panel.}
\label{table:1}
\begin{tabular}{l | c c c c c c c c c }
\hline
%  & \tablehead{1}{l}{b}{Phase\\}
%  & \tablehead{1}{c}{b}{Time\\Myr}
%  & \tablehead{1}{c}{b}{ M$_1$\\$\mso$}
%  & \tablehead{1}{c}{b}{ M$_2$\\$\mso$}
%  & \tablehead{1}{c}{b}{ P\\d}
%  & \tablehead{1}{c}{b}{ $\varv_{\rm rot,2}$\\km~s$^{-1}$}  
%  & \tablehead{1}{c}{b}{ Y$_{\rm c,2}$\\}
%  & \tablehead{1}{c}{b}{ Y$_{\rm s,2}$\\}
%  & \tablehead{1}{c}{b}{ $\varv_{\rm orbit,2}$\\km~s$^{-1}$} \\
 Phase   \T       & Time & M$_1$  & M$_2$ &  P  & 
$\varv_{\rm rot,1}$  & $\varv_{\rm rot,2}$ & Y$_{\rm c,2}$ & Y$_{\rm s,2}$ & $\varv_{\rm orbit,2}$  \\   
  \T       & Myr & $\mso$ & $\mso$ & d &
km~s$^{-1}$ & km~s$^{-1}$ &  &  & km~s$^{-1}$ \\

\hline
1 ZAMS               & 0      & 16     & 15    & 5.0    &  230    & 230 & 0.248 & 0.248   & 201    \\            
2 begin Case B       & 9.89   & 15.92  & 14.94 & 5.1    &   96    & 85  & 0.879 & 0.248   & 198    \\
3 end Case B         & 9.90   &  3.93  & 20.77 & 38.2   &   27    & 719 & 0.434 & 0.348   & 29     \\
4 ECCB primary       & 11.30  &  3.71  & 20.86 & 42.7   &   40    & 767 & 0.457 & 0.441   & 27     \\
                     &        &        &       &        &         &     &       &         &        \\ 
5 ECHB secondary     & 18.10  & --     & 16.76 & --     &    --   & 202 & 0.996 & 0.956   & --     \\
6 ICB  secondary     & 18.56  & --     & 12.85 & --     &    --   & 191 & 0.000 & 0.996   & --     \\      
7 ECCB secondary     & 18.56  & --     & 12.83 & --     &    --   & 258 & 0.000 & 0.996   & --     \\
\hline
\end{tabular}
\end{table}

\section{Results}

The evolution of the binary system proceeded as follows (cf. Table 1). The initial rotational velocity
of both stars has been set to $230\kms$, but both stars synchronize with the orbital
rotation within about 1\,Myr, to equatorial rotational velocities
of only about $50\kms$. Rotationally induced mixing before
the onset of mass transfer is thus negligible --- in contrast to typical O~stars evolving
in isolation \citep{2000ApJ...544.1016H,2000A&A...361..101M}. The initially more massive star ends core hydrogen burning
after $\sim 9.89\,$Myr, and Case~B mass transfer begins shortly thereafter. It sheds
about $12\mso$ evolving into a $\sim 4\mso$ helium star. About 1.5\,Myr later, it sheds another
$\sim 0.2\mso$ as a helium giant, before exploding as Type~Ib/c supernova.

The mass gainer keeps about $6\mso$ of the overflowing matter, rendering the mass accretion
efficiency to roughly 50\%. Thereafter, it enters a phase of close-to-critical rotation,
which induces rejuvenation and quasi-chemically homogeneous evolution (Fig.~\ref{plots}).
 Its mass loss is enhanced by rotation. About 5\,Myr after the
onset of accretion, the surface helium mass fraction of the mass gainer is increased to
values above 60\%, and Wolf-Rayet mass loss is assumed from then on. The star finishes core
hydrogen burning after another 3\,Myr, at an age of 18.1\,Myr, with a mass of 16.8$\mso$,
a surface helium mass fraction of 95\%, and rotating with $\sim 200\kms$ at the surface. 

After core hydrogen exhaustion, the mass gainer contracts and spins-up to critical rotation,
which leads to a mass shedding of almost 2$\mso$.  
During its remaining lifetime of less than 0.5\,Myr, it loses about another 2$\mso$
to a Wolf-Rayet wind. It ends its life as a rapidly rotating Wolf-Rayet star
with a final mass of about $13\mso$, ready to form a collapsar (average specific angular momentum in the CO core $<j_{\rm CO}>\simeq 2\times10^{16}$ cm$^2$s$^{-1}$). 
Assuming the binary brakes up due to the explosion of the mass loser, the mass gainer would
have traveled for about 7\,Myr with its final orbital velocity of 27$\kms$
a distance of about 200\,pc.

%Following \citet{1995MNRAS.274..461B} we calculate the kick velocity 
%necessary to unbind the binary system, under the hypothesis of instantaneous removal of the SN ejecta. 
%Two extreme values correspond to the most and to the least efficient 
%geometrical configuration for the supernova kick to break up the system. 
%The minimum kick velocity is $52\kms$ and corresponds to the case where the kick is aligned with the orbital 
%velocity vector of the supernova progenitor. The maximum kick velocity necessary to unbind the system 
%is $350\kms$, which is required if the kick is aligned to the orbital velocity vector, but directed backward.
%According to the observed velocity distribution of radio pulsars, about 55\% of pulsars have a space velocity
%larger than $350\kms$, while more than 98\% have a velocity above $52\kms$ \citep{2002ApJ...568..289A}.
%In order to  estimate the chance of obtaining a runaway star out of our system we performed a Monte Carlo simulation 
%for a randomly oriented supernova kick. 
%According to the observed velocity distribution of radio pulsars \citep{2002ApJ...568..289A} the probability for the 
%binary system to break up by the first supernova explosion is about 80\%. 

%\paragraph{<A subsubsubsection>}

\section{Discussion}
The binary evolution model presented above shows that quasi-chemically 
homogeneous evolution may  occur in mass gainers of low-metallicity massive 
early Case~B binaries. The comparison of the mass gainer with a corresponding single star model made it clear that
such binary components evolve in the same way as extremely rapidly rotating
single stars. This confirms that the scenario of quasi-chemically homogeneous evolution
might not be restricted to single stars, but may apply to the accreting component of
massive close binaries as well.

While we provide only one example, it seems likely that this scenario
applies to most massive close binary components which accrete or gain an appreciable
amount of mass; this may encompass Case~A binaries and early Case~B binaries
\citep{1992ApJ...391..246P,1999A&A...350..148W,2001A&A...369..939W}.
Case~A merger are also likely contributing to this scenario.
While the merged object will have more mass than the initially more massive star
in the binary, the product will be extremely rapidly rotating due to the 
orbital angular momentum, as in the case of some blue stragglers
\citep{1993ASPC...53....3L}.

\subsection{Binaries and the distribution of rotational velocities}

The best constraint so far on the distribution of initial rotational velocities (IRF)
comes from the recent study of young O~stars in the SMC, mostly from the cluster
NGC~346 \citep{2006A&A...456.1131M}. According to \citet{2006A&A...460..199Y}, the three most rapid
rotators from the sample of 21 O~stars would qualify for the quasi-chemically
homogeneous evolution scenario, and remarkably, all three stars are found to be
helium-enhanced. The simplest approach to understand those stars is to assume that 
they correspond to the tip of the IRF.

However, that data of \citet{2006A&A...456.1131M} reveals another interesting feature:
two of the the three mentioned stars are runaway stars, with radial velocities
deviating by 30...70$\kms$ from the average cluster radial velocity. While dealing with
low number statistics, this information opens another possibility:
that the most rapidly rotating young O~stars in the SMC are products of binary evolution.
A closer examination of the IRF derived by \citet{2006A&A...456.1131M} appears to support this
idea: While the three rapid rotators show $\varv\sin i \simgr 290\kms$, all other
stars have $\varv\sin i \simle 210\kms$.  

The following hypothesis therefore seems conceivable: The IRF of single O~stars in the SMC
ends at about $210\kms$ --- too early to allow quasi-chemically homogeneous evolution
and collapsar formation. However, massive close binary evolution enhances the IRF to 
what we may call the apparent IRF as measured by \citet{2006A&A...456.1131M},
which leads to the redshift dependent GRB rate as worked out by \citet{2006A&A...460..199Y}.
According to the binary population synthesis model of \citet{1992ApJ...391..246P},
about 10\% of all massive binaries might lead to a Case~A merger or early Case~B
mass transfer, which is sufficient to populate the rapidly rotating part of
the IRF of Mokiem et al. 
In that context, the rapidly rotating O~star in the sample of \citet{2006A&A...456.1131M} which
does not appear as runaway star could either have an undetected high proper motion,
or it could be the result of a Case~A merger --- where no runaway is produced.

\subsection{Effects from runaway GRBs}

The runaway nature of a GRB progenitor, as obtained in our example,
has important observational consequences for
both the positions of GRBs, and their afterglow properties.
Concerning the afterglow, it is relevant that
the medium close to a WR star has the density profile of a free-streaming wind, and analytical 
and numerical calculations both suggest that the free wind of a single WR star typically extends over 
many parsec \citep{2006A&A...460..105V}. 
However, from the analysis of GRB afterglows, a constant circumstellar medium density has been 
inferred in many cases \citep{2000ApJ...536..195C,2001ApJ...554..667P,2002ApJ...571..779P,2004ApJ...606..369C}.
A possible explanation has been proposed by \citet{2006A&A...460..105V}, who simulated the circumstellar 
medium around a moving WR star.
As the GRB jet axis is likely perpendicular to the space velocity vector, 
the jet escapes through a region
of the bow-shock where the wind termination shock is very close to the star.
Therefore, the jet may enter a constant density medium quickly in this situation.

Concerning the GRB positions, since the spin axis of the stars in a close binary system are likely orthogonal to the
orbital plane, the observation of a GRB produced by the proposed binary channel is possible only 
if the binary orbit is seen nearly face on.
Then the direction of motion of the runaway GRB progenitor must be orthogonal to the line of sight, 
allowing the progenitor, for the given space velocity, to obtain the maximum possible 
apparent separation from its formation region.
The finding of \citet{2006A&A...454..103H},
that the nearest three long gamma-ray bursts may be due to runaway stars is in remarkable
agreement with our scenario. While the collapsar progenitor in our binary
model travels only 200~pc before it dies, compared to the 400...800~pc deduced by \citet{2006A&A...454..103H},
binary evolution resulting in higher runaway velocities are certainly possible \citep{2005A&A...435.1013P}. 
It remains to be analyzed whether the runaway scenario is compatible with the finding that
long GRBs are more concentrated in the brightest regions of their host galaxies than core collapse
supernovae \citep{2006Natur.441..463F}.

%%%%%%%%%%%%%%%%%%%%%%%%%%%%%%%%%%%%%%%%%%%%
%% Sample figure:
%%
%% The option [height=...] scales the picture to the given height,
%% without it it would be printed at its nominal size
%%%%%%%%%%%%%%%%%%%%%%%%%%%%%%%%%%%%%%%%%%%%

%\begin{figure}
%  \includegraphics[height=.3\textheight]{golfer}
%  \caption{Picture to fixed height}
%\end{figure}

%%%%%%%%%%%%%%%%%%%%%%%%%%%%%%%%%%%%%%%%%%%%%%%%
%% BACKMATTER
%%%%%%%%%%%%%%%%%%%%%%%%%%%%%%%%%%%%%%%%%%%%%%%%

%\begin{theacknowledgments}
  
%\end{theacknowledgments}


%\bibliographystyle{mn2e}      % Modified 
\bibliographystyle{aipproc}   % if natbib is available

\bibliography{7115ref}

\IfFileExists{\jobname.bbl}{}
 {\typeout{}
  \typeout{******************************************}
  \typeout{** Please run "bibtex \jobname" to optain}
  \typeout{** the bibliography and then re-run LaTeX}
  \typeout{** twice to fix the references!}
  \typeout{******************************************}
  \typeout{}
 }

\end{document}